\documentclass{article}
\usepackage{hyperref}
\usepackage{amscd,amsmath,amsthm,amssymb}
\usepackage{enumerate,varioref}
\usepackage{epsfig}
\usepackage{graphicx}
\usepackage{mathtools}
\usepackage{tikz}
\usepackage{multirow}
\usepackage{xcolor,colortbl}
\newtheorem{thm}{Theorem}

\theoremstyle{definition}
\newtheorem{defn}[thm]{Definition}
\theoremstyle{remark}
\newtheorem{ex}[thm]{Example}
\newtheorem{rem}[thm]{Remark}
\numberwithin{equation}{section}

\newcommand{\R}{\mathbb{R}}
\newcommand{\Z}{\mathbb{Z}}

\newcommand{\half}{\frac{1}{2}}
\newcommand{\ket}[1]{|#1\rangle}

\newcommand{\bracket}[2]{\langle #1 | #2 \rangle }

\title{Using Quantum Mechanics to Cluster Time Series}
\author{Nousot Applied Research Group \footnote{Corresponding Author: Clark Alexander. Co-Authors  Luke Shi, Sofya Akhmametyeva. NARG also includes  Rami Jachi, Mahyar Moghadam, Ziyi Liu, Zeynep Abat}\\ email \href{mailto:clark@nousot.com}{clark@nousot.com}}

\begin{document}
\maketitle

\begin{abstract}
In this article we present a method by which we can reduce a time series into a single point in $\R^{13}$.  We have chosen 13 dimensions so as to prevent too many points from being labeled as ``noise." A quick calculation reveals that in high dimensions, nearly all points are more than three standard deviations away from the center in multiple aspects.  When using a Euclidean (or Mahalanobis) metric, a simple clustering algorithm will with near certainty label the majority of points as ``noise."  On pure physical considerations, this is not possible.  Included in our 13 dimensions are four parameters which describe the coefficients of a cubic polynomial attached to a Gaussian picking up a general trend, four parameters picking up periodicity in a time series, two each for amplitude of a wave and period of a wave, and the final five report the ``leftover" noise of the detrended and aperiodic time series. Of the final five parameters, four are the centralized probabilistic moments, and the final for the relative size of the series.  The first main contribution of this work is to apply a theorem of quantum mechanics about the completeness of the solutions to the quantum harmonic oscillator on $L^2(\R)$ to estimating trends in time series.  The second main contribution is the method of fitting parameters.  After many numerical trials, we realized that methods such a Newton-Rhaphson and Levenberg-Marquardt converge extremely fast if the initial guess is good.  Thus we guessed many initial points in our parameter space and computed only a few iterations, a technique common in Keogh's work on time series clustering. Finally, we have produced a model which gives incredibly accurate results quickly.  We ackowledge that there are faster methods as well of more accurate methods, but this work shows that we can still increase computation speed with little, if any, cost to accuracy in the sense of data clustering.
\end{abstract}

\tableofcontents

\section{Introduction}

This work began in the summer of 2017 with the task of building a new clustering algorithm.  Obviously, there are dozens of clustering algorithms which serve different purposes.  There are even qualitative measures as to overall performance of each algorithm.  In the case of spatial and categorical data, there are many techniques, algorithms, methods, etc to compute roughly any particular thing a researcher or data scientist could wish for.  So then, what makes one want to write a new algorithm?  In the case of NARG, we belong to a research group which has built its first product on time-series predictions.  Again, there are already many techniques available for predicting time series.  The point for NARG is to cut down the initial search space by clustering time series and comparing against the representative samples so as do dramatically speed up the search and reduce computation time.  In our solution to this problem we solved the problem of time series clustering in what we believe to be not a wholly new method, but a new method which combines existing methods in a unique way.\\

This article is structured in the following way. We begin our discussion with known problems for clustering time-series and we give some of the widely known results around this.  Most of them are due to Keogh and his colleagues\cite{Keogh1,Keogh2,Keogh3,Keogh4}.  Some of his ideas were instrumental in providing us with insights as to increased speed. We continue our discussion of clustering by realizing that while function spaces have many distance metrics, they are not all equivalent.  This is a little statement in that Euclidean space has several metrics which are equivalent.  They do not all give the same distances, but many are bound by constant multiples of others, and under the equivalence relation $\Theta$ on real valued functions, we declare Euclidean metrics equivalent.  No such class exists for functions.  Even in the simple case of functions on a finite interval we immediately run into the problem of which basis to choose.  Different projections into finite dimensional subspaces alter the measurements greatly.  We will discuss some of the well known distances in function spaces.  We also mention the idea of translating functions and changing phases and how this affects calculations of distances between two functions.\\

In section four we begin exploring the slightly unusual constraint of 13 dimensions.  The basic idea is that a multivariate Gaussian spreads out too much in Euclidean space when we use too many dimensions.  We appeal to Neumann's idea of fitting an elephant with four parameters and try this three times.  The thirteenth parameter calculates how much in total is ``leftover." In many engineering applications this will be called the power spectral density.  After we have subtracted away the major trends and the periodic trends, we should have a very tiny leftover piece.  If, however, we have a series which is so noisy that we can't pick off a trend with a Gaussian or several cosines, then we will report a large ``noise" component which will shift the final five dimensions away and thus we have achieved a great amount of information as we claim within the constraints of a relatively small number of dimensions.\\

Section five will be a simple review of the quantum harmonic oscillator in one dimension.  We need not go into great specifics about homological properties of the Weyl algebra, but we will get to one main theorem that we apply in this work.  Namely
\begin{thm}
Let $D = \frac{d}{dx}$ and consider the linear differential operator:
\[
H = \half\left(x^2 - D^2 \right)
\]
$H$ has spectrum $\{ (n + 1/2) | n\in \Z, n\ge 0\}$ with eigenfunctions $\ket{n} = H_n(x)e^{-x^2/2}$ where $H_n(x)$ are the Hermite polynomials. That is
\[
\forall n\ge 0, \; H\ket{n} = \left(n+\half\right)\ket{n}
\]

These functions $\ket{n}$ form a complete orthonormal basis for $L^2(\R)$.  In particular, these functions form a basis for probability density functions supported on the real line.
\end{thm}

Given this information we proceed in section six by showing how to fit parameters to solutions of the quantum harmonic oscillator in one dimension in order to pick off major trends in time series.  We show that rather than considering the time series as a wave itself, we consider the time series as a non-normalized pdf.  We tried fitting these as wave functions, but the computational cost of fitting complex parameters is too expensive. One of our main objectives, however, is to reduce computational cost without losing too much accuracy.

In order to achieve this, we shift the entire time series up or down so that its minimum value is zero.  In this way we can now think of our series as a pdf.  We then appeal to the Weierstrass transform to show that we can consider fitting the parameters;
\[
f(x;\alpha) = (\alpha_0 + \alpha_1 x + \alpha_2 x^2 + \alpha_3 x^3)e^{-x^2/2}
\]
Rather than having to fit
\[
f(x,\alpha) = \alpha_0\ket{0} + \alpha_1\ket{1} + \alpha_2\ket{2} +\alpha_3\ket{3} 
\]

These are equivalent in their fitting even if the values of the parameters would be somewhat different.\\

Sections two through six introduce our main motivation for using the quantum harmonic oscillator.  We see in section six that a linear combination of the ground state and first three eigenstates of the harmonic oscillator provide us a robust framework for fitting many types of time series.  In section seven we turn to actually fitting the parameters.  The discussion herein begins with Newton's method for root finding in one-dimension and we extend this to multiple dimensions.  We define the function we wish to minimize, which is a sum of squared errors from the actual data to the oscillator.  From here we consider the time points as positions and we minimize in the parameter space.  As we see in Newton's method for root finding we can extend this to an optimization method by finding roots of the gradient.  However, this is terribly computationally expensive, especially considering that we will need to compute a second derivative of a sum of functions.  The Levenberg-Marquardt algorithm speeds things up dramatically.  In essence, we replace the many curvature components in the second derivatives by straight lines.  This has one advantage and one disadvantage.  The disadvantage is that we don't fit our parameters by moving as close as possible at each iteration.  This, after all, is what Newton's method does for us; it follows the tangent plane to the plane of height zero.  In the optimization method, we follow the best paraboloids to the plane of height zero.  Levenberg-Marquardt replaces some of the curved pieces by straight lines.  This is analogous to planning a hiking route by looking at a 2D map without regard to the level curves on a mountain.  The major advantage in this method is speed.  If our guess is ``reasonably close" then the iterative guess is an excellent guess. We additionally speed up this process by recognizing that Levenberg-Marquardt converges very quickly, and so we only compute three iterations for each starting guess.  This takes very little time for each guess.  We split up the parameter space into a several initial guesses.  Essentially, we guess each parameter to begin as a positive or negative or sometimes zero value and run the algorithm.  Since three iterations can be computed extraordinarily fast, we allow ourselves many initial guesses. After each pass we collect the best overall approximation.  Whichever approximation gets the closest we keep and throw away all the rest.  Thus we are able to fit this with an in-place algorithm.\\

In section eight we subtract off the model with the best fitted parameters then try again with a periodic model.  Then we subtract off the periodic model and we are left with what we called the noisy or detrended data.\\

In section nine we give some estimates about how much noise is leftover.  In essence we report the first four central moments of the detrended data.  We appeal again to the method of generalized moments which tells us that the first four moments are sufficient to capture the vast majority of information.  In many circumstances the detrended data will no longer have mean zero with unit standard deviation.  Skewness is an important component in the leftover noise as it gives us some clues as to whether the data is becoming more noisy or less.  Finally, the kurtosis tells us how wide the tails are.  One should expect the tails are wider after the data has been detrended.  If the resultant kurtosis is small, then our model has picked up a sufficient amount of the information.  Finally, we report the overall size of the noise.  This gives us some idea of how long the time series is.  If two time series produce the first twelve parameters identically (this will happen with probability zero) and the last is uneven, then the larger thirteenth parameter tell us that this is a longer time series.\\

In section ten we present some numerical results, including our time trials for dimensionality reduction using several well-known techniques including PCA, ISOMAP, and t-SNE.  Our clustering results are given for Federal Reserve Economic Data (FRED).  We see that our algorithm produces excellent results with excellent computational speed.  The results we show test against a time series package called Cesium.  As our results show, in some cases our reduction occurs at ten times the speed.\\

We conclude by giving a high level overview of the algorithm we have used.  In future work we would like to optimize this work by testing another distance measure.  The next distance for time series we would like to work through is dynamic time warping.

\section{The Trouble with Clustering Time Series}

Let's begin with an easy example.  Consider the following four time series all of length 50.

\[
\includegraphics[scale = 0.48]{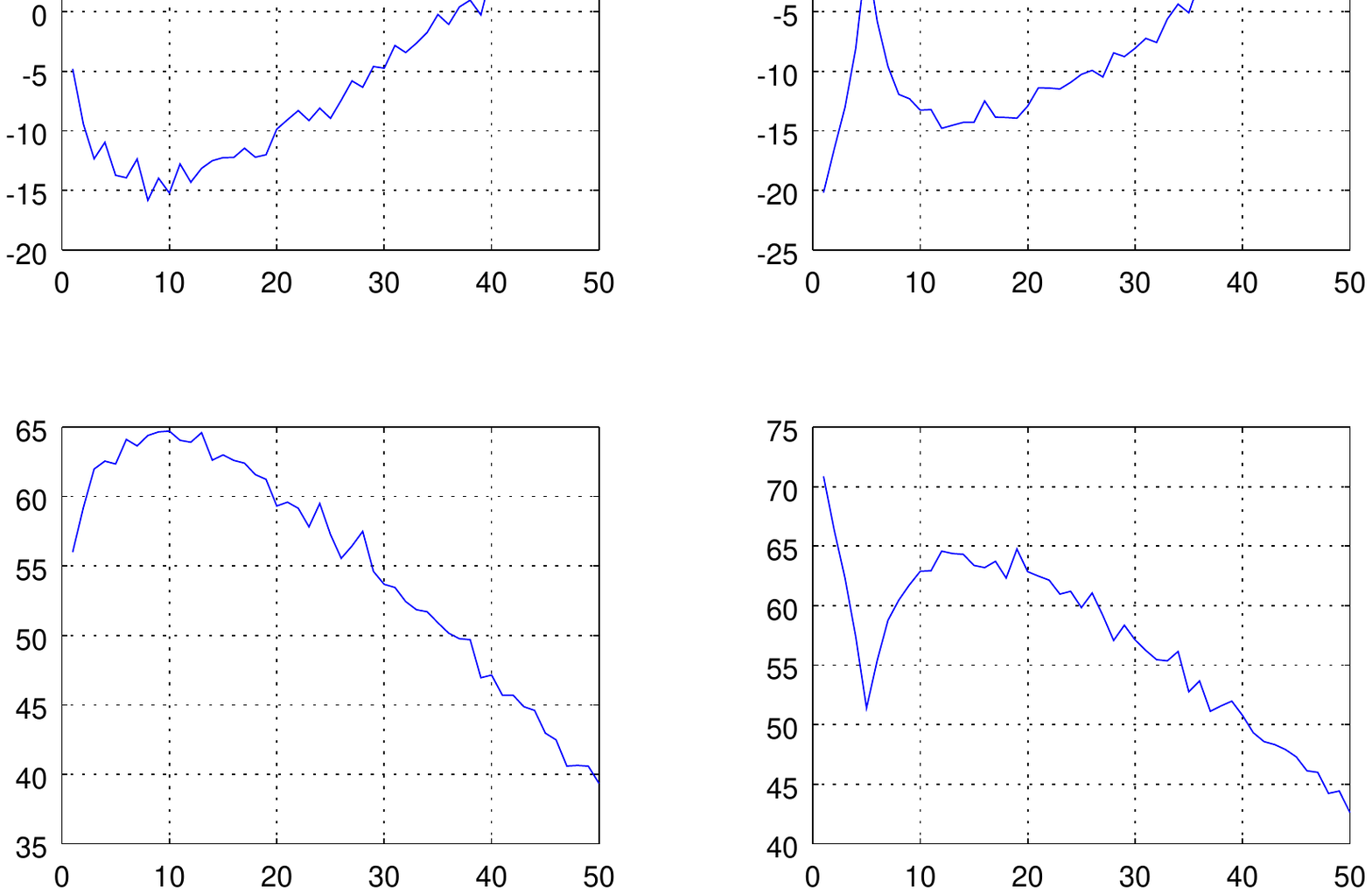}
\]

These are all the same time series with some minor shift.  The upper left time series is the function
\[
f(t) = t - 10*\ln(|t|+1) + 0.65*w_t
\]
where $w_t$ is a normally distributed random variable.

The upper right is simply a shift $f(t-5)$.  The bottom row gives us $-f(t)$ and $-f(t-5)$.  In many cases these will be viewed as incredibly different time series.  However, if we consider the example of two coffee shops in mid-sized cities in the United States, say Kansas City and Oklahoma City, opening 6 months apart, they will likely have similar sales in early months, but one shop will look as if it's much further ahead of the other if we consider the monthly sales for each month on its own (i.e. August sales in OKC vs August sales in KC) rather than month 1 sales vs month 1 sales.  These time series have the same relation as graphs 1 and 2 or graphs 3 and 4.

In terms of regression analysis, one often considers minimizing the sum of squares of vertical distances.  Even for relatively small differences in translations (left or right) this can result in enormous distances between two functions which are otherwise identical.  One such technique that addresses this problem is dynamic time warping, hereafter DTW.  The goal of DTW is to make the distance between two time series as small as possible by considering the closest points on each time series.  Essentially this allows us to translate functions left or right a little bit without affecting the overall distance much.  In our research we found DTW to be very effective at minimizing these translations.  However, in terms of computational complexity, computing the DTW distance between two time series of lengths $t_1$ and $t_2$ respectively, we require $O(t_1t_2)$ steps.  When one time series is particularly long, this is not feasible on a large set of time series in which one needs to compute all the mutual DTW distances, in fact, if we have $n$ such series all of a length relatively equal to $t$ we require $O(n^2 t^2)$ steps to complete.  This is completely unfeasible when $n>100,000$ and $t\approx 100$.  Again, Keogh et al.\cite{Keogh3,Keogh4} have several techniques to cut down the computation time.  An interesting task is to combine Keogh's technique of filtering by a lower bound and considering each time series as a probability distribution function.\\

We again reach the classical problem, how much accuracy can we trade for speed?  Simple regression methods are fast, but often inaccurate without a sufficiently good guess.  DTW is accurate, but computationally expensive.  Even in the best case scenario, where we can cut off our computations at a low threshold we still require $\Omega(n^2)$ comparisons at least, to account for all time series.  This assumes we only have to look at one point in each pair of time series.  Of course, this is completely unrealistic.  Hopefully, we will have reduced our computation from $O(n^2 t^2)$ to somewhere in the range $O(n^2 t)$.

\section{Distances in Function Space and Their Computation}

In mathematical analysis we often define a distance by a metric.  A metric, of course, is a function
\[
d: X \rightarrow \R_+
\]

with the following properties
\begin{enumerate}
\item $d(x,x)\ge0$\\
\item $d(x,y) = d(y,x)$\\
\item $d(x,y)+d(y,z)\ge d(x,z)$.
\end{enumerate}

A metric space then is a pair $(X,d)$ and we usually write
\[
d(x,y) = \|x-y\|
\]

Then we can define the ``size" or ``distance" of a point $x\in X$ by 
\[
\|x\| = d(x,0)
\] 

In Euclidean space, when we have an inner product
\[
\langle , \rangle : \R^n \times \R^n \rightarrow \R
\]

we can define distance the standard way:
\[
\|x\| = \sqrt{\langle x,x\rangle}
\]
which tells us 
\[
d(x,y) = \|x-y\| = \sqrt{\langle x-y,x-y\rangle}.
\]

None of this is particularly interesting on its own, or especially in finite dimensional Euclidean space.  Our problem is quite a bit larger: the set $X$ we wish to metrize is a \emph{function space} which means we need to find some method by which we can put a distance on functions.\\

In function spaces, we have many considerations.  First of all is a measure.  In the case of this work, we wish to approximate time series by smooth functions.  Time series are in no way smooth, but we will try to build a curve $f(x)$ whose points at the measured times are ``as close as possible" to the sequence $(a_j)_{j=1}^{t}$ which are the values of the time series.\\

For the moment let's simply take the measure to be the Lebesgue measure $\mu$ so that we may define two common types of metrics on the function space $C^{\infty}(\R)$ (smooth functions of one real variable).  The first common family are the $L^p$ spaces.  We define them as such.  Let $f\in C^{\infty}(\R)$ then for $0<p<\infty$ its $p$-norm is defined by
\[
\|f\|_p = \left(\int_{\R} |f|^p d\mu\right)^{1/p}
\]

The higher the value of $p$ the more importance the maximum value of $f$ gets.  In particular
\[
\|f\|_{\infty} := \max_{\R}(|f|)
\]

These $L^p(\R)$ spaces are dual to $L^q(\R)$ spaces in the sense that if $f\in L^p$ and $g\in L^q$ then we can compute their inner product when 
\[
\frac{1}{p}+\frac{1}{q} = 1.
\]  
However, we immediately see that when $p\ne 2$ there is no guarantee that we can take an inner product of $f$ with itself.  Thus we often prefer to work with $L^2(\R)$.  This is the only self-dual space, and thus an inner-product space.  As we will see the quantum harmonic oscillator provides an orthonormal basis of $L^2(\R)$ and so $L^2$ becomes a Hilbert space.  None of this is particularly shocking to anyone who has had the scantest amount of undergraduate analysis, but it is a useful discussion for us since we debated for quite some time the ``best" way to measure a function before we could cluster time series.\\

The second somewhat common family of metrics are the Sobolev norms.  In our case, we will only consider the Sobolev norms on $L^2(\R)$ since we already see the necessity is using inner products.  Given a function $f\in C^k(\R)$ we define its $k$-Sobolev norm by
\[
\|f\|_{k,2} = \left(\sum_{j=1}^k \|D^j f\|_2\right)^{\half}
\]

That is, we take the inner product ($L^2$) of the vector
\[
(f,Df,D^2f,\dots,D^kf)
\]

where $D^jf  = \frac{d^j f}{dx^j}$

This is useful to because it allows us, by integration by parts, to compute the inner product of a smooth function against a function which is not smooth.  In our case, we see that computing the inner product of a time series against a smooth function now makes sense.  In this sense, we realize that a time series, while not smooth, can be approximated arbitrarily closely by a finite sequence of smooth functions \cite{Evans}. 

Thus it makes sense in a meaningful way to use a sequence of smooth $L^2$ functions to approximate a time series.  We find an appropriate approximating set in \S5. 

\section{Why Thirteen Dimensions?}

Perhaps instead of giving the computation directly, we should take a moment to reflect on the idea of distances in high dimensions.  In Euclidean space the distance (analogous to $L^2$) between vectors $x$ and $y$ is
\[
\langle x-y,x-y \rangle = \|x\|^2 - 2\langle x,y\rangle + \|y\|^2
\]

The higher the number of dimensions, the less likely that randomly chosen vectors $x$ and $y$ will be ``close" to each other.  Thus we expect $2\langle x,y \rangle$ to be relatively small versus the contributions of $\|x\|^2$ and $\|y\|^2$.  In data clustering, very high dimensional sets tend to have a vast majority of points marked as ``noise" and with good reason.  Even if we consider the Mahalanobis metric (Euclidean metric on $z$ scores) many values will be between 
$-3$ and $3$.  That is, we don't often see values outside of 3 standard deviations.  In fact, in one dimension, strictly less than $1/9$ of values can be outside of 3 standard deviations.  This is in the absolute worst case. (cf. Chebyshev's rule).  Of course, with a large number of samples, we expect the ensemble to converge to Gaussian (or a $t$-distribution with high degrees of freedom) wherein greater than 99\% of values are within the three standard deviation bound.  However, when we add more dimensions the likelihood of every variable being within 3 standard deviations in every dimension falls to $(1-\varepsilon)^n$.  That is, in $n$ dimensions things get noisy.
\[
(1-\varepsilon)^n \approx 1 - n\varepsilon + \binom{n}{2}\varepsilon^2
\]

This is how we arrived at the idea of calculating the number of dimensions we should use.  Suppose our distribution guarantees just a bit more than 99\% of values within three standard deviations, possibly not the estimate 99.73\% is in a true standard normal, but close.  How many dimensions do we need to keep the vast majority of points from being ``noisy?"  For our considerations of accuracy versus overall speed of computaiton we settled on requiring the size of the noisy points to be less than $1/8$ the total sample.  We arrived at
\[
(.99)^{13} \approx \frac{7}{8}
\]

So we decided that 13 dimensions would keep enough of the points to give us useful information even in simple clustering algorithms.  We now embark on how we turn a time series into a point in 13 real dimensions.\\

An important remark is that, one can certainly use fewer dimensions, but the models are less accurate.  In the end, we find 13 dimensions to be an excellent balance between speed and accuracy for our purposes.  The package for Python called Cesium computes far more dimensions and is incredibly more accurate, but as we shall see in the section on results and time trials (\S10), is close to an order of magnitude slower.

\section{The Quantum Harmonic Oscillator and its Solutions}

Without rehashing too much of what is already known about the quantum harmonic oscillator let's simply state the problem, give its solutions and explain why we know that they will help us in time-series clustering.  The Hamiltonian for a quantum harmonic oscillator with mass $m=1$ and frequency $\omega$ is given by
\[
H = \half\left(p^2 + (x\omega)^2\right)
\]
Here we have used the ``natural" units $\hbar = 1$.  Now we also know that in one dimension the quantum momentum operator is 
\[
p = -i\frac{d}{dx} = -iD
\]

Thus we rewrite our Hamiltonian simply as
\[
H = \half\left((\omega x)^2-D^2\right)
\]

This suggests factoring the operator into a difference of squares.  Unfortunately, for us, position and momentum do not commute as we have the relation
\[
[D,x]=1
\]

\begin{rem}
The commutation $[D,x]=1$ is simultaneously the Leibniz rule for derivatives in first term calculus, as well as the rudimentary equation from which we derive Heisenberg's uncertainty principle.
\end{rem}

In our attempt to factor the Hamiltonian we define two new operators $a$ and $a^*$

\begin{eqnarray}
a & = & \sqrt{\frac{\omega}{2}}\left(x + \frac{D}{\omega}\right) \\
a^* & = &  \sqrt{\frac{\omega}{2}}\left(x - \frac{D}{\omega}\right) 
\end{eqnarray}

Since we can compute 
\[
[a,a^*] = 1
\]

We can rewrite the quantum harmonic oscillator in an algebraic form rather than a differential equation:
\[
H = \omega \left(a^* a + \frac{1}{2}\right)
\]

Giving us the easy ground state solution:
\[
a\psi_0 = 0 \implies \psi_0 = A e^{-\omega x^2/2}
\]

where $A$ is a normalization constant.  From here we use the fundamental relation $[a,a^*]=1$ to arrive at a full set of solutions: 
\[
\psi_n = A_n (a^*)^n \psi_0
\]

Where $A_n$ are the normalization constants.  These functions are commonly written as
\[
\ket{n} = \psi_n = A_n H_n(\sqrt{\omega} x)e^{-\omega x^2/2}
\]

Again, $A_n$ are the normalization constants and $H_n(z)$ are the physicists' Hermite polynomials.

\begin{rem}
There are differences in the physicists' Hermite polynomials and the probabilists' Hermite polynomials.  Basically, since the functions $\ket{n}$ are not thought of as probability distributions themselves, but rather as wave functions, which are square roots of pdfs with an additional phase, the physicists' Hermite polynomials are not monic, but the probabilists' polynomials are.
\end{rem}

However, the Hermite polynomials are all polynomials of degree $n$.  For example, in the physcists' setting
\begin{eqnarray}
H_2(z) & = & 4z^2 - 2\nonumber\\
H_3(z) & = & 8z^3 - 12z\nonumber\\
H_4(z) & = & 16z^4 -48 z^2 + 12\nonumber
\end{eqnarray}

Now comes the major theorem that we make use of for clustering time series:

\begin{thm}
The set of functions $\{\ket{n}\}_{n=0}^{\infty}$ forms a complete orthonormal basis for functions in $L^2(\R)$.
\end{thm}

The proof of this theorem is rather straightforward and can be found in any graduate analysis or probability book.  The skeleton of the proof goes something like this: The functions are orthonormal as we can show by considering
\begin{equation}
\bracket{n}{m} = \int_{\R} H_n(x)H_m(x) e^{-x^2} dx = \delta_{n,m}
\end{equation}

Notice that in the integral the weight is $e^{-x^2}$ since we are multiplying two copies of $e^{-x^2/2}$, one for each wave function.  Again, this is why physicists and probabilists have different versions of Hermite polynomials.

After we have established orthonormality, we show completeness by showing that 
\[
\forall f\in L^2(\R),\ket{n} \; \bracket{f}{n} = 0 \implies f\equiv 0 (\text{almost everywhere})
\]

This is done by recognizing that
\[
H_{2n}(x) \text{ is even.}
\]
\[
H_{2n+1}(x) \text{ is odd.}
\]

Then splitting $f$ into its even and odd parts:
\[
f = f_e + f_o
\]

If $n$ is even then
\[
\bracket{f_e}{n} \ge 0 
\]
since the integral is everywhere nonnegative. Thus $f_e \equiv 0$.  If $n$ is odd then
\[
\bracket{f_o}{n} \ge 0 
\]

Since each integral 
\begin{eqnarray}
\int_{-\infty}^{0} f_o H_{n}(x)e^{-x^2/2}dx &=&0\nonumber\\
\int_0^{\infty} f_o H_{n}(x)e^{-x^2/2}dx &=&0\nonumber
\end{eqnarray}

Since each half is an even function.  Thus $f_o \equiv 0$.
Since $f = f_e+f_o \equiv 0 + 0 \equiv 0$ $f$ is zero almost everywhere.  And the solutions to the harmonic oscillator form a complete orthonormal basis on $L^2(\R)$.

\section{Turning a Time Series into a  Wave Function}

Bolstered by a pair of classical theorems, we now approach turning our time series into wave functions.  Actually, we will turn our time series into non-normalized probability distribution functions.  Our theorems, of course, are the first from PDEs which says any Sobolev function can be approximated by a convergent sequence of smooth functions.  The second is that for every function which is square integrable on the real line, we can approximate it via solutions of the quantum harmonic oscillator.\\

Now we turn to several more theorems or techniques so that we can approximate our time series as quantum harmonic oscillators.  We recognize that pdfs require non-negative values.  The means the first thing we do is to shift the data up or down so that the minimum value is actually zero.  Strictly speaking, this is not necessary, it is entirely possible to approximate finite time series with negative values by using harmonic oscillator solutions.  Our next question becomes, how many energy levels should we use?  That is, in the basis $\ket{n}$ we can write any function $f\in L^2(\R)$ as
\[
f = \sum_{n=0}^{\infty} \alpha_n\ket{n}
\]

However, it is of no practical use to approximate a finite time series with an infinite sequence.  This, in fact, does exactly the opposite of what we desire, making things more complicated and more expensive to compute rather than less.  Thus we must ask the question: how many energy levels should we use to get a ``good approximation"?  What $N$ should we choose so that
\[
f \approx \sum_{n=0}^{N} \alpha_n \ket{n} ?
\]

From \S4 we expect that $N\le 12$.\\

The next theorem we use is that for a given probability distribution function, if the moment generating function (mgf) exists then the pdf can be determined uniquely almost everywhere from the mgf. In our assumptions we are using smooth functions so our mgf will uniquely determine our time series approximation.\\

By the central limit theorem, a sequence of sums of independent random variables converges to a standard normal.  This suggests to us that we need to capture at least three pieces of information, the amplitude (since we do not require our shifted time series to be normalized), the mean, and standard deviation.  The $n^{th}$ moment of a probability distribution $f$ is
\[
\mu_n = \int_{\R} (x-\mu)^n f(x) dx
\]
And in our case $\mu = 0$ since we have a Gaussian.  So we expect to be able to fit at least $\mu_0,\mu_1,\mu_2$ suggesting we should try to fit a function of the form
\[
f(x) = (\alpha_0 + \alpha_1 x + \alpha_2 x^2)e^{-\omega x^2/2}
\]

This is basically an autoregressive model (AR(2)) attached to a Gaussian.  However, we require a broad and robust model.   By solving general autoregressive models with two and three lags (second and third order recurrence relations) we see that AR(3) models allow us a much broader class of functions than an AR(2) model.  Thus we elect to fit one more moment and we look for functions of the form
\begin{equation}
f(x) = (\alpha_0 + \alpha_1 x + \alpha_2 x^2+\alpha_3 x^3)e^{-\omega x^2/2}
\end{equation}

Our next natural question is whether or not we can transform this into a wave function:
\[
\sum_{n=0}^{3} \beta_n \ket{n}
\]

\subsection{Weierstrass Transforms}

In this section we explain in two ways why fitting parameters to a wave function is equivalent to fitting 

\[
f(x;\alpha) = (\alpha_0 + \alpha_1 x + \alpha_2 x^2+\alpha_3 x^3)e^{-\omega x^2/2}
\]

The first method is to take the Weierstrasss transform of $f$.  

\begin{defn}
Given a smooth function $f\in L^2(\R)$ we define the Weierstrass transform $Wf$ by
\[
W_tf(y) = \frac{1}{\sqrt{4\pi t}}\int_{\R} f(x) e^{\frac{-(x-y)^2}{4t}}dx
\]
\end{defn}

We have several important properties of Weierstrass transforms which we will use.  The transforms $W_t$ form a semigroup in the parameter $t$.
\[
W_{t + s} = W_t \circ W_s
\]

This allows us to manipulate the frequency $\omega$ in our wave function.  We will take advantage of this fact when we begin fitting our parameters.\\

The standard Weierstrass transform, however, is one in which $t=1$.  In this way, we can formally write:
\begin{equation}
Wf(x) = e^{D^2}f(x) 
\end{equation}

If we expand this as a Taylor series, we see that it picks up all the even probabilistic moments.  However, we should not overlook the fact that many distributions do not have moment generating functions, because some finite moment becomes infinite.  To our advantage the functions that we are trying to fit decay rapidly, that is they are $L^2(\R)$, Schwartz functions.  Thus we can formally apply The Weierstrass transform in the form $e^{D^2}$.  In image processing, one might recognize this as a so called ``low-pass filter."  The particularly important result for us is the following:
\begin{equation}
W[x^n] = H_n\left(\frac{x}{2}\right)
\end{equation}

That is, the Hermite polynomials and functions $x^n$ are, up to scaling, Weierstrass and inverse Weierstrass transforms of each other.\\

This lands us on the first justification for why we can simply use functions of the form
\[
f(x;\alpha) = (\alpha_0+\alpha_1x+\alpha_2 x^2+\alpha_3 x^3)e^{-\omega x^2/2}
\]

The second idea we use is that we can simply expand our first few Hermite polynomials, collect coefficients and realize that
\[
\alpha_0 \ket{0} + \alpha_1 \ket{1} + \alpha_2 \ket{2} + \alpha_3 \ket{3} = (\beta_0 + \beta_1 x+\beta_2 x^2+\beta_3 x^3) e^{-\omega x^2/2}
\]

\section{Fitting the Parameters}

In this section, we need to decide how to fit the four parameters $\alpha_0,\dots,\alpha_3$ so as to ``best" determine the trends of a given time series.  For this work we have chosen a variant on nonlinear least squares regression.  In this particular case we define $\hat{y}_k$ the actual measurement at time $k$ and our \emph{residues} are

\begin{equation}
r_k = \hat{y}_k - f(k;\alpha)
\end{equation}

Then we take the function distance (technically an $\ell^2$ norm) as the sum of squared errors:
\begin{equation}
S(\alpha) = \sum_{k=1}^{t} r_k^2 = \sum_{k=1}^{t} (\hat{y}_k-f(k;\alpha))^2
\end{equation}

Where $t$ is the length of the time series.  We should also mention that our data is sequential (often weekly or monthly) with cleaned data so we take $k=1$ to be the time at which the first measurement was made.  We can certainly change variables to consider the sequence of events $k=2000$ to $t+ 1999$.  This only shifts the time axis, not the actual values, so we start with $k=1$.  

\begin{rem}
We start with $k=1$ because most of the parameter fitting was done in Octave.  In Python and many other languages, one would start at $k=0$.  Again, this simply shifts the time axis, not the values which we are trying to minimize.
\end{rem}

In this situation we're claiming to fit four parameters, but part of our idea in using the quantum harmonic oscillator is that we have an additional hidden parameter $\omega$ that gives us some idea of energy.  Let's take a look at this parameter.\\

A standard Gaussian distribution already carries with it two parameters:
\[
e^{\frac{-(x-\mu)^2}{2\sigma^2}}
\]

We will set these $\mu$ and $\sigma$ to center our quantum state at the center of the time series and to have the entire time series cover 6 ``standard deviations."\\

Thus our algorithm begins: Given $\{y_k\}_{k=1}^t$ a time series
\begin{enumerate}
\item[1.] $t$ is the length of the time series. So we physically center the data replacing our function:
\[
f(x;\alpha) = \sum_{j=0}^{3} \alpha_j x^j e^{-{x^2}/2\sigma^2}
\]

By the more physical
\begin{equation}
f(x;\alpha) = \sum_{j=0}^{3} \alpha_j \left(x - \left(\frac{t}{2}\right)\right)^j e^{-(x-t/2)/2\sigma^2}
\end{equation}

\item[2.] We want our fitted function to cover $6\sigma$ over the space of $t$. See \S4 for the justification of this. Since we have centered our time series at $t/2$, we look $3\sigma$ left and $3\sigma$ to the right.
\[
t = 6\sigma \implies \sigma = \frac{t}{6} \implies \frac{1}{2\sigma^2} = \frac{18}{t^2}
\]

Thus we wish to fit the major trends of the time series with the quantum state.
\begin{equation}
f(x;\alpha) = \sum_{j=0}^{3} \alpha_j \left(x- \left(\frac{t}{2}\right)\right)^j e^{-18(x-(t/2))^2/t^2}
\end{equation}

Bringing this into the physical interpretation:
\begin{equation}
\frac{\omega}{2} = \frac{18}{t^2} \implies \omega = \frac{36}{t^2} = \frac{1}{\sigma^2}
\end{equation}

Our ground state energy is therefore inversely proportional to variance.  This is equivalent to the fact that wavelength is directly proportional to variance. A fact that will be well known to physicists, but perhaps not as well recognized in machine learning circles.

\item[3.] In order to make the time series into a non-normalized probability distribution we shift the values $\{\hat{y}_k\}$ to
\[
m = \min\{\hat{y}_k\} \rightarrow \hat{y}_k := \hat{y}_k - m.
\]

\end{enumerate}

Now we wish to fit our four $\alpha$ paramters.  The analysis which follows is a modification of the Levenberg-Marquardt method:\\

We begin by Newton's root finding method.  In one dimension we wish to solve an equation $f(x^*) = 0$.  Pick an initial guess $x_0$ then follow the tangent line at $(x_0,f(x_0))$ to the axis to reveal our new guess $x_1$ which gives us an iterative procedure.

\begin{equation}
x_{n+1} = x_n - \frac{f(x_n)}{f'(x_n)}
\end{equation}  

We need to acknowledge that this gives no guarantee to solve $f(x^*) = 0$.  There are multiple ways in which this fails.  For example, $f$ may not have a root, we may have an initial guess which lands on a local extremum which makes the next guess undefined. However, when Newton's method works, it works extremely well, and converges quadratically fast.  One more problem we must acknowledge, especially in this context, is that $f$ may have multiple roots, and so the iterates $x_n$ may become chaotic as in the Mandelbrot and Julia sets.  So we must take extra care not to waste computational effort on iterating chaotic guesses.\\

We extend Newton's method for root finding to multiple dimensions by:

\begin{equation}
x_{n+1} = x_n - (DF(x_n))^{-1}F(x_n)
\end{equation}

In the case that $DF$ is not square we can consider the pseudo-inverse.\\

We are, in fact, looking for a root, but it's not the root of the function $S(\alpha)$.  Rather we are looking to minimize $S(\alpha)$ which says that we're looking for the zero gradient; that is, we wish to find the root of
\[
F(\alpha) = \nabla_{\alpha} S(\alpha)
\]

In order to find an extremum we'll use Newton's method on the gradient.  Thus we wish to iterate over
\[
\vec{\alpha}_{n+1} = \vec{\alpha}_{n} - DF(\vec{\alpha}_n)^{-1} F(\vec{\alpha}_n)
\]

Notice that we have defined 
\begin{equation}
F(\alpha) = \begin{bmatrix}
\partial S / \partial\alpha_0 \\
\partial S / \partial\alpha_1 \\
\partial S / \partial\alpha_2 \\
\partial S / \partial\alpha_3 
\end{bmatrix}
\end{equation}

which gives us the somewhat horrific looking formula:
\begin{equation}
DF = \begin{bmatrix}
\partial^2 S / (\partial\alpha_0)^2 & \partial^2 S / \partial\alpha_0 \partial \alpha_1 & \partial^2 S / \partial\alpha_0 \partial \alpha_2 & \partial^2 S / \partial\alpha_0 \partial \alpha_3 \\
\partial^2 S / \partial\alpha_1 \partial \alpha_0 & \partial^2 S / (\partial\alpha_1)^2  & \partial^2 S / \partial\alpha_1 \partial \alpha_2 & \partial^2 S / \partial\alpha_1 \partial \alpha_3 \\
\partial^2 S / \partial\alpha_2 \partial\alpha_0 & \partial^2 S / \partial\alpha_2 \partial \alpha_1 & \partial^2 S / (\partial\alpha_2)^2 & \partial^2 S / \partial\alpha_2 \partial \alpha_3 \\
\partial^2 S / \partial\alpha_3 \partial\alpha_0 & \partial^2 S / \partial\alpha_3 \partial \alpha_1 & \partial^2 S / \partial\alpha_3 \partial \alpha_2 & \partial^2 S / (\partial\alpha_3  )^2
\end{bmatrix}
\end{equation}

The situation becomes much worse when we realize that $S$ is already a sum of residues.  Taking a closer look:
\[
S = \sum_{k=1}^{t} (\hat{y}_k - f(k;\alpha))^2
\]

This tells us:
\begin{equation}
F = \nabla S = \sum_{k=1}^{t} 2(\hat{y}_k - f(k;\alpha))(-\nabla f(k;\alpha))
\end{equation}

Additionally, we have:

\begin{equation}
DF = \nabla(\nabla S) = 2\sum_{k=1}^{t} (-\nabla f(k;\alpha)) (-\nabla f(k;\alpha))^T + (\hat{y}_k - f(k;\alpha))(-\nabla(\nabla f(k;\alpha))) 
\end{equation}

The obvious problem for solving this numerically is, of course, all the cross terms in the second parametric derivatives in the wave function.\\ 

The Levenberg-Marquardt algorithm replaces the matrix $\nabla(\nabla f)$ with a diagonal matrix.  The reasoning for making this substitution is two-fold.  First, if the residues are ``small" then the matrix $DF$ is approximately
\[
DF \approx 2\sum_{k=1}^{t} (\nabla f)(\nabla f)^T
\]

This is a sum of singular matrices of rank 1.  There is no guarantee that this matrix is invertible.  Reasoning back to Newton's method in one dimension, one of the potential problems we see is when the function in question doesn't have a root, but the minimum is barely positive, e.g., $x^2+\varepsilon$.  This gets close to the minimum quickly, but the iterated guess diverges afterward.  When we add a full rank matrix to a singular matrix, then result is invertible with high probability.  In the case of actual data with noise, the resultant matrix is invertible with probability 1.  Even so, we can consider the pseudo-inverse in which we invert the nonzero eigenvalues.\\

The second reason is that the second derivatives are computationally expensive.  Replacing with a diagonal matrix is inexpensive.  However, the replacement of the second derivatives with a diagonal matrix inherently declares
\[
\frac{\partial^2 f}{\partial \alpha_i \partial \alpha_j} = 0 \; \text{ if } i\ne j
\]

Geometrically this gets rid of our curvatures.  The result is that our algorithm fits a little bit less accurately, but much faster.\\

In order to take advantage of our speed while recognizing the limitations of Newton's method we run this curve-fitting scheme multiple times.  Recalling that one of the potential failures of Newton's method is that we are not guaranteed to find a global minimum, but rather some local minimum.  If we're given a convex function, then we find a root, if one exists.  In this situation, we do not have a convex function, but rather the derivative of a convex function.  We do, however, guarantee that there is a global minimum.  By the way we have calculated our distances, we can never get a value for $S < 0$.\\

Thus our current setup is that we are guaranteed the existence of a global minimum, with no idea at all of the number of local maxima or minima in existence.  Furthermore, we're fitting our parameters to give smaller sums by assuming all the cross second derivatives are zero.\\

We give no guarantee of finding the true global minimum without exponentially much effort, thus we must rely on the fact that our algorithm runs quickly with the simplified numerics, and if our initial guess is ``good" then our parameters will converge to a local minimum quickly.  This certainly happens if many of the residues are small.\\

Our best strategy therefore is to compute very few iterates with very many initial guesses and then take the overall best result.

Consider the following example: We have a level set 
\[
\cos(x^2 + 2x - y - y^2) = \half
\]
 
\[
\includegraphics[scale=0.5]{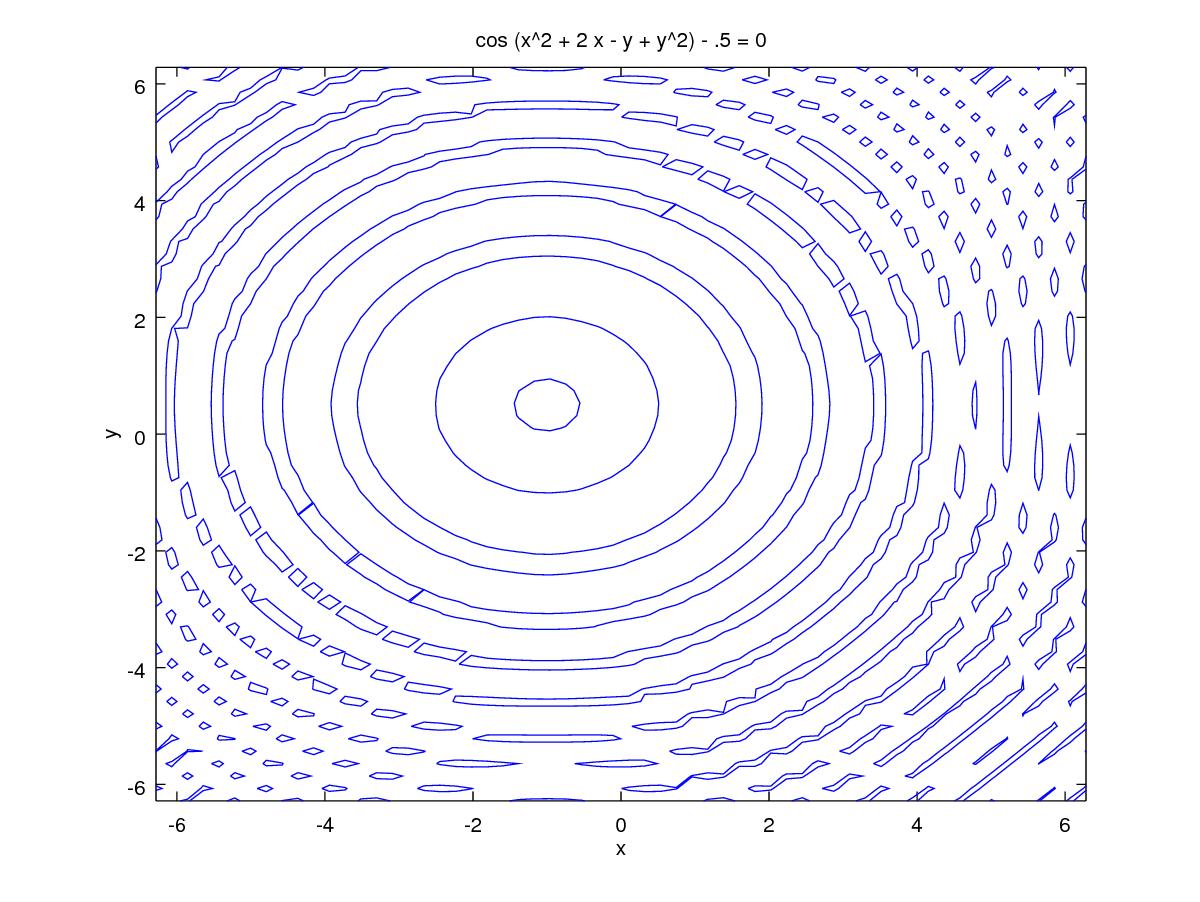} 
\]

Clearly there are infinitely many local minima here.  Choosing a particular initial guess for Newton's method will most likely land us on the closest local minimum or maximum.  However, we have absolutely no guarantee of attaining the correct global minimum.  For example, one possible trajectory of Newtonian guesses looks like:

\[
\includegraphics[scale=0.5]{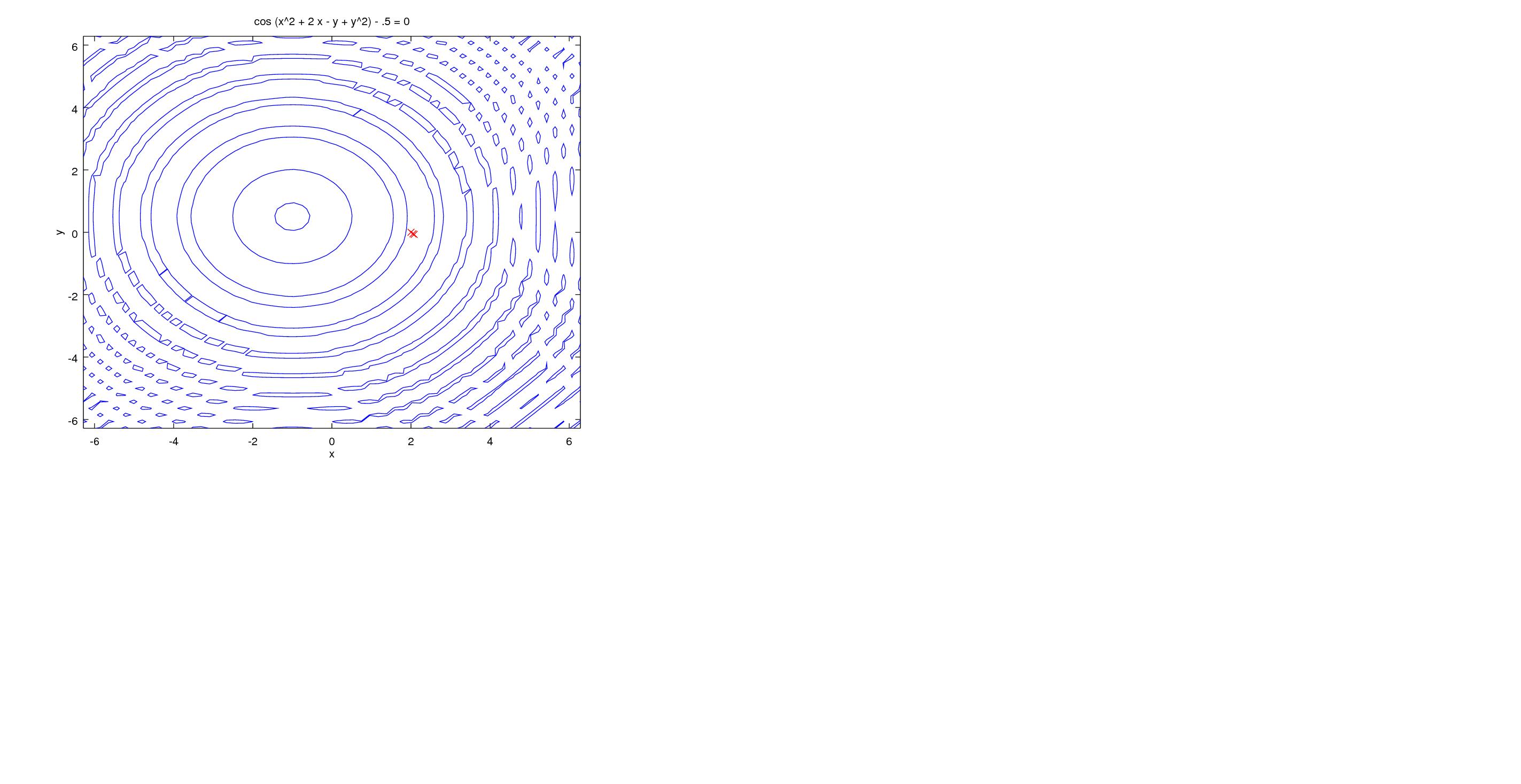} 
\]
where the red `x' markings follow a trajectory beginning at (2,0).

For an initial guess which is close to the solution, both methods, Newton and Levenberg-Marquardt,
%Newton's method and the Levenberg-Marquardt algorithm
follow trajectories which are very close, but the latter computes its iterations much more quickly.  In order to give ourselves the best chance to find an approximate global minimum we divide our space as such:
\[
\includegraphics[scale=0.5]{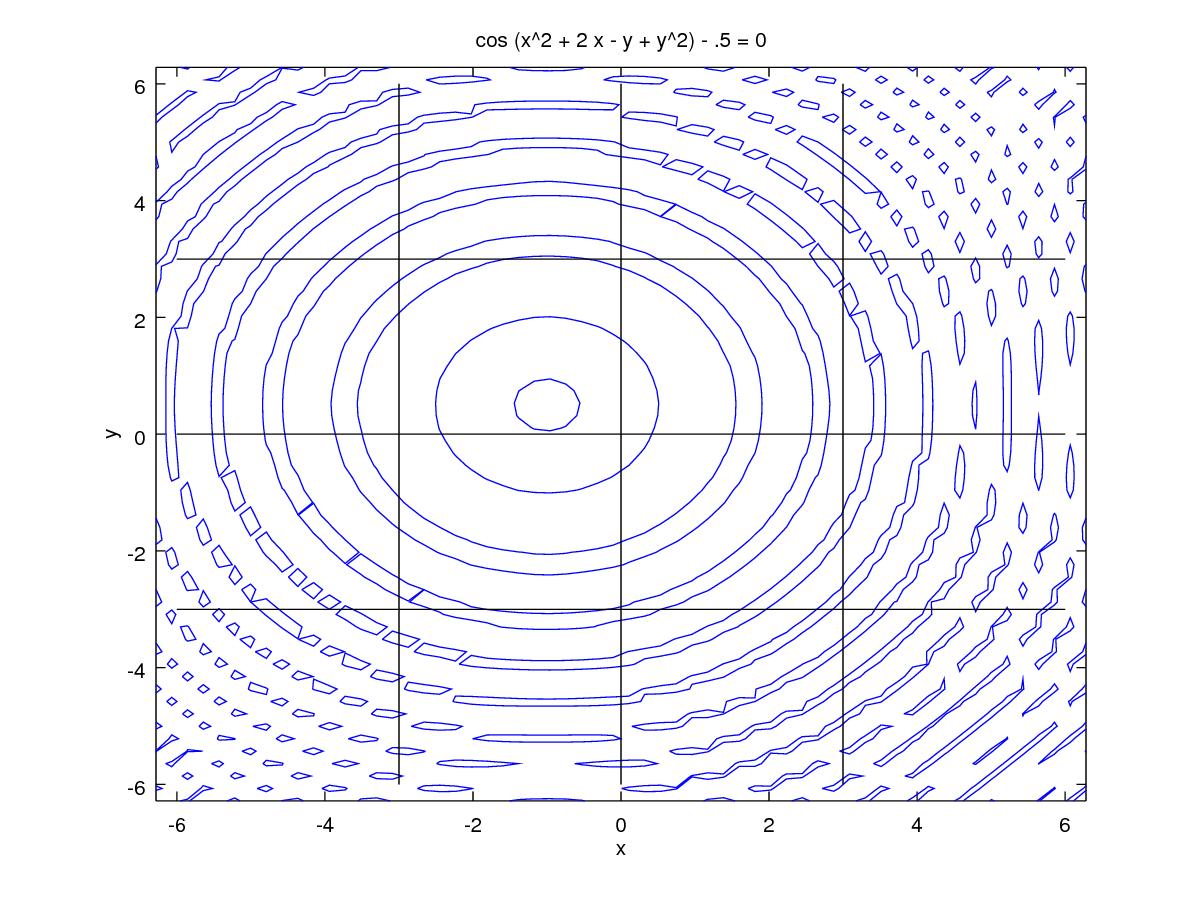} 
\]

In each cell we pick an initial starting point and then iterate three times in using the Levenberg Marquardt method.  This dramatically increases our chances of finding the global minimum while only increasing our computation time by a fixed constant the number of initial guesses.  In a particularly cumbersome time series we may sort our best final iterations and find that a few of them drastically outperform the others.  We can then refine our search by reducing our search domain.
\[
\includegraphics[scale=0.5]{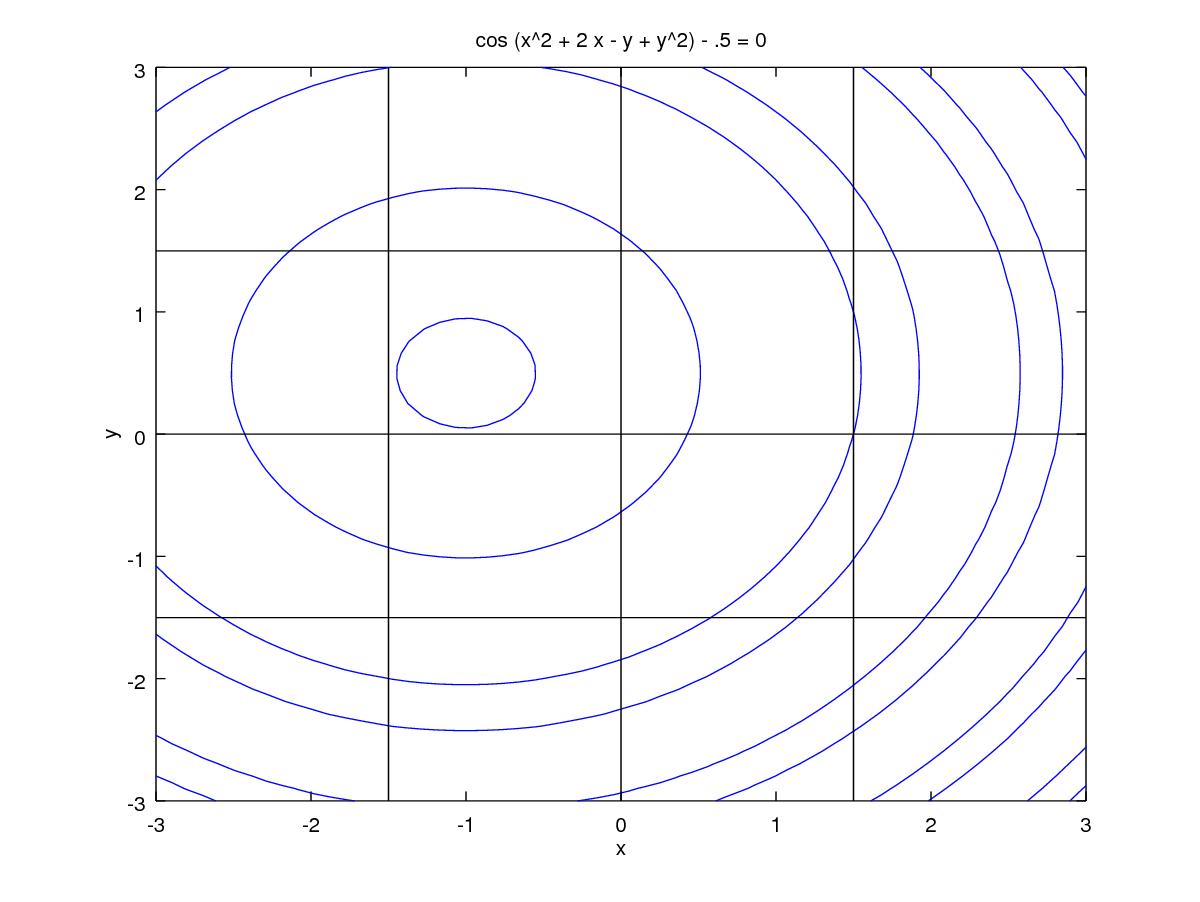} 
\]

Again, we only increase our computation time by a constant multiple.  As we have it drawn we check 16 initial guesses. Then we cut down our search space and try 16 more intial guesses.  This gives us only a multiple of 32 times we check our guesses, each of which is computationally fast.  In time trials on the principal author's local machine, an off-the-shelf laptop, we tested Newton's method for the full model in 13 dimensions.  Fitting one time series with high accuracy took roughly 4 and a half minutes.  After the computational reductions we were able to compute 36 per minute.  This is an increase in speed of nearly 450 times.  There was a slight decrease in accuracy, but it did not affect the overall clustering.  We recognize that this means the decrease in accuracy was less than the noise threshold for our clustering algorithms.  Thus in computational terms we increased our speed by two and a half orders of magnitude while losing no accuracy at the level of clustering.

\section{Detrending the Data}

At this point most of the hard work has been done.  We have selected an extraordinarily robust model in the quantum harmonic oscillator to pick up the overall trends in time series and we have chosen a method which is much faster given a single initial guess to iterate multiple times over multiple initial guesses.  We have thus picked out the ``best trend" we can with the speed we desire.\\

To now we have simply shifted the data, centered the data, and picked off the trend.  At this point we wish to subtract off the trend and model the data for periodicity.  That is we consider the following transformation of our data.
\begin{enumerate}
\item Original data ${d_i}$ of length $t$.\\
\item $d_i$(new) $= d_i - \min_i(d_i)$\\
\item $f(\alpha_*)$ is the fitted trend: $d_i$(detrended) $ = d_i$(new) $-f(\alpha_*)$
\end{enumerate}

From here we perform the same algorithm, but rather than fitting an oscillator, we fit a truncated Fourier series:
\[
g(\beta) = \beta_0 \cos(\beta_1 x) + \beta_2\cos(\beta_3 x).
\]

We fit the detrended data with a periodic function again by a modified Levenberg-Marquardt method. This leaves us with
\[
d_i(\text{aperiodic}) = d_i(\text{detrended}) - g(\beta_*)
\]

For the sake of clustering this is roughly as much as we need to do for the sake of trends and periodicities.  Obviously we can pick out more pieces of information, but the more we report, the more spread out our coordinates become and the harder it is for a clustering algorithm, especially a distance-based clustering algorithm, to classify our points as anything but noise.  This is a problem we ran into using Cesium.  More on this in \S10.

\section{How Much Noise is Left?}

As we initially $z$-scored our data, our expectation is that the mean is now zero and the variance is now unity.  However, after having subtracted the trends and periodicity, our mean may no longer be zero.  This is common is very noisy data.  Our method only gives a smooth approximation to the time series and thus if a lot of noise is leftover, the mean may have shifted.  Additionally, we hope the variance has been decreased a little.  However, when considering the time series as a probability distribution, we expect some sort of skewness.  The quantum harmonic oscillator can pick up some skewness, especially in the terms $\ket{1}$ and $\ket{3}$, but heavily skewed graphs will still maintain a large amount of skewness.  Finally, we consider how wide the distribution is by calculating the centered kurtosis of the detrended and aperiodic data. Thus we report the first four centered probabilistic moments of the detrended and aperiodic data.  Hopefully this will be relatively little.  Those who have done any amount of time series analysis will, however, recognize that modeling the noise is the ``interesting" part of the analysis.  Since the oscillator is robust enough to pick up a wide variety of trends and we then consider periodicity, we will often have only the noise leftover.  For ``nice" time series this will be almost zero, for very lengthy time series this will be somewhat larger.  For a final coordinate we report the spectral power density, which is the sum of magnitudes squared of the Fourier coefficients.  However, by Parseval's equation we know this is equal to the sum of magnitudes squared of the data itself.  Again in the interest of speed, we already have real valued data and rather than requiring the computer to compute a Fourier transform, deal with complex numbers, and reconvert, we simply report:
\[
\sum_{i=1}^{t} d_i(\text{aperiodic})^2
\]

For very long time series, this is a good coordinate to distinguish between two time series of similar trends and periodicities.

\section{Time Trials and Results} 

In this section we present some results of two methods.  The first is a feature selection package called cesium\cite{Cesium}.  It is the belief of the authors that this package was originally developed for applied astronomy or computational molecular chemistry.  This package looks through a time series and reports 38 features, including number of local minima, maximum value, minimum value, length, period, minor periods, etc.  It provides a highly accurate description of a time series.  NARG had good results with its information output.  Some of the code was modified to fit our needs, but the modifications were only to prevent compiler errors in a different version of Python.  The second method we compare is our in-house clustering method.  The data set we use is the publicly available Federal Reserve Economic Data (FRED).  These data sets are labeled only by number, not by title.  For example, we do not know which FRED data corresponds to unemployment, nor to inflation rates.  There are approximately 200 data sets here, each of which was turned into a spatial point via both methods.  For the purposes of visualization we compared several dimensionality reduction methods and preprocessing methods. Below are the results of the dimensionality reductions to two and three dimensions via five algorithms and three different preprocessing methods.  The algorithms were run  on the same machine and thus the results yield an accurate cross-comparison.  In the first two columns of each table we have raw data, that is, the coordinates as they come from the time series.  The second pair of columns performs a standardization of zero mean and unit variance in each coordinate. The third pair of columns involves preprocessing the coordinates by applying the distribution function of the $\Gamma$ distribution with mean $\mu = k\theta$ and standard deviation $\sigma = \sqrt{k}\theta$.  This is an in-house technique that we developed at Nousot, as we found preprocessing columns by the Gamma distribution gives a very  clear delineation of clusters, even in an algorithm as simple as k-means clustering.   In more sophisticated clustering methods preprocessing by the Gamma distribution yields excellent results.\\

Each of these algorithms takes roughly 200 time series, and produces a spatial point.  The preprocessing is obviously faster with fewer dimensions, when it comes to standardization or distributing via Gamma.  We had also attempted preprocessing by the Rayleigh distribution, Boltzmann distribution, and the exponential distribution.  Each gave interesting results.  The authors and NARG in general observed the best results with standardization ($z$-scoring) and Gamma distributing.  Since we have equally sized datasets the dimensionality reduction is a good test for clustering.  Once the data are reduced to two or three dimensions, the final clustering will require the same number of operations, which will translate into the same amount of time.  Thus, we present here the results of dimensionality reduction in lieu of the clustering which we performed later.  The first table shows the results of reducing to two dimensions.  We see that in raw data, Cesium performs better on average, although the difference in Multidimensional scaling (MDS) is small enough that this might be considered a draw.  Isomap also presents three very close results.  Principal component analysis, however, is where NARG shines through.  The results are roughly an order of magnitude faster.\\

However, we must realize that using two dimensions is a rather blunt tool.  We prefer a bit more nuance.  In each table we report the time to completion to the nearest $10^{-5}$ seconds

%Raw
%Cesium
%t-SNE
%mds
%spectral
%isomap
%pca
%2D Embedding 
%7.386579990386963
%0.14578008651733398
%0.041342973709106445
%0.05490398406982422
%0.0022668838501
%3D Embedding
%15.909281969070435
%0.1634068489074707
%0.03391003608703613
%0.04991888999938965
%0.00165104866028

%raw
%narg
%t-SNE
%mds
%spectral
%isomap
%pca
%2D Embedding 
%7.226201057434082
%0.14719390869140625
%0.0519258975982666
%0.050637006759643555
%0.000596046447754
%3D Embedding
%17.25080394744873
%0.1571810245513916
%0.030475854873657227
%0.04752206802368164
%0.000569105148315

%Standard
%Cesium
%t-SNE
%MDS
%SPECTRAL
%ISOMAP
%PCA
%2D Embedding 
%7.335690021514893
%0.13344192504882812
%0.02423095703125
%0.0662379264831543
%0.00317907333374
%3D Embedding
%17.010974884033203 
%0.14614415168762207
%0.04020094871520996
%0.06579303741455078
%0.0019109249115

%Standardized
%narg
%t-SNE
%MDS
%SPECTRAL
%ISOMAP
%PCA
%2D Embedding 
%6.706430912017822 
%0.13163995742797852
%0.02686285972595215
%0.062261104583740234
%0.000655889511108
%3D Embedding
%16.961781978607178 
%0.14674997329711914
%0.021337032318115234
%0.05845999717712402
%0.000633001327515

%Gamma
%Cesium
%t-SNE
%MDS
%SPECTRAL
%ISOMAP
%PCA
%2D Embedding 
%9.169102907180786
%0.13297605514526367
%0.018301963806152344
%0.04487299919128418
%0.00219011306763
%3D Embedding
%20.283581018447876
%0.14696407318115234
%0.020895957946777344
%0.05270195007324219
%0.00146389007568

%Gamma
%narg
%t-SNE
%MDS
%SPECTRAL
%ISOMAP
%PCA
%2D Embedding 
%7.129094123840332
%0.13507604598999023
%0.05990314483642578
%0.032653093338012695
%0.000530004501343
%3D Embedding
%17.60053300857544
%0.1510450839996338
%0.025274038314819336
%0.03200197219848633
%0.00052809715271

\begin{table}[ht]
\caption{Projections into 2 Dimensions}
\begin{center}
\begin{tabular}{| c | c | c | c | c | c | c|}
\hline
\multirow{2}{*}{Method $\backslash$ Analysis} & \multicolumn{2}{c|}{Raw Data} & \multicolumn{2}{c|}{Standardized} & \multicolumn{2}{c|}{$\Gamma(k,\theta)$}\\

  &Cesium & NARG & Cesium & NARG &  Cesium &  NARG\\
\hline
&&\cellcolor{cyan}&&\cellcolor{cyan}&&\cellcolor{cyan}\\
t-SNE & 7.38658 &\cellcolor{cyan} 7.22620 & 7.33569 & \cellcolor{cyan}6.70643 & 9.16910 & \cellcolor{cyan}7.12909\\  
&&\cellcolor{cyan}&&\cellcolor{cyan}&&\cellcolor{cyan}\\
\hline
&\cellcolor{cyan}&&&\cellcolor{cyan}&\cellcolor{cyan}&\\
Multidimensional Scaling & \cellcolor{cyan}0.14578 & 0.14719 & 0.13344 & \cellcolor{cyan}0.13164 & \cellcolor{cyan}0.13298 & 0.13508\\  
&\cellcolor{cyan}&&&\cellcolor{cyan}&\cellcolor{cyan}&\\
\hline
&\cellcolor{cyan}&&\cellcolor{cyan}&&\cellcolor{cyan}&\\
Spectral & \cellcolor{cyan}0.04134 & 0.05192 &\cellcolor{cyan} 0.02423 & 0.02686 &\cellcolor{cyan} 0.01830 & 0.05990 \\  
&\cellcolor{cyan}&&\cellcolor{cyan}&&\cellcolor{cyan}&\\
\hline
&&\cellcolor{cyan}&&\cellcolor{cyan}&&\cellcolor{cyan}\\
Isomap & 0.05490 & \cellcolor{cyan}0.05064 & 0.06624 & \cellcolor{cyan}0.06226 & 0.04487 &\cellcolor{cyan} 0.03265\\  
&&\cellcolor{cyan}&&\cellcolor{cyan}&&\cellcolor{cyan}\\
\hline
&&\cellcolor{cyan}&&\cellcolor{cyan}&&\cellcolor{cyan}\\
PCA & 0.00227 & \cellcolor{cyan}0.00060 & 0.00318 &  \cellcolor{cyan}0.00066 & 0.00219 &\cellcolor{cyan} 0.00053\\  
&&\cellcolor{cyan}&&\cellcolor{cyan}&&\cellcolor{cyan}\\
\hline
\end{tabular}
\end{center}
\label{tab:multicol}
\end{table}

Below are the results for reducing to three dimensions.  In this case we see NARG's results begin to pull away steadily.  Interestingly these results improve as we consider ever more dimensions.  For example, it requires zero time to reduce NARG's results to 13 dimensions, while is requires a bit of work to reduce larger dimensions. Let's look at some notable highlights in three dimensions.  NARG continues to outperform handily via PCA, increases its lead in Isomap. Where Cesium outperfomed handily in spectral dimensionality reduction to two dimensions, in three dimensions NARG is slightly faster on raw data and nearly double speed in the standardization.  The t-statistic nearest neighbor embedding (t-SNE) is significantly slower in every metric, but again under the three dimensional Gamma distribution preprocessing, NARG outperforms by roughly 15\%.

\begin{table}[ht]
\caption{Projections into 3 Dimensions}
\begin{center}
\begin{tabular}{| c | c | c | c | c | c | c|}
\hline
\multirow{2}{*}{Method $\backslash$ Analysis} & \multicolumn{2}{c|}{Raw Data} & \multicolumn{2}{c|}{Standardized} & \multicolumn{2}{c|}{$\Gamma(k,\theta)$}\\

  &Cesium & NARG & Cesium & NARG &  Cesium &  NARG\\
\hline
&\cellcolor{cyan}&&&\cellcolor{cyan}&&\cellcolor{cyan}\\
t-SNE & \cellcolor{cyan}15.90928 & 17.25080 & 17.01097 & \cellcolor{cyan}16.96178 & 20.28358 &\cellcolor{cyan} 17.60053\\  
&\cellcolor{cyan}&&&\cellcolor{cyan}&&\cellcolor{cyan}\\
\hline
&&\cellcolor{cyan}&\cellcolor{cyan}&&\cellcolor{cyan}&\\
Multidimensional Scaling & 0.16341 &\cellcolor{cyan} 0.15718 &\cellcolor{cyan} 0.14614 & 0.14675 &\cellcolor{cyan} 0.14696 & 0.15104\\  
&&\cellcolor{cyan}&\cellcolor{cyan}&&\cellcolor{cyan}&\\
\hline
&&\cellcolor{cyan}&&\cellcolor{cyan}&\cellcolor{cyan}&\\
Spectral & 0.03391 &\cellcolor{cyan} 0.03048 & 0.04020 & \cellcolor{cyan}0.02134 &\cellcolor{cyan} 0.02090 & 0.02527\\  
&&\cellcolor{cyan}&&\cellcolor{cyan}&\cellcolor{cyan}&\\
\hline
&&\cellcolor{cyan}&&\cellcolor{cyan}&&\cellcolor{cyan}\\
Isomap & 0.04992 &\cellcolor{cyan} 0.04752 & 0.06579 & \cellcolor{cyan}0.05846 & 0.05270 &\cellcolor{cyan} 0.03200\\  
&&\cellcolor{cyan}&&\cellcolor{cyan}&&\cellcolor{cyan}\\
\hline
&&\cellcolor{cyan}&&\cellcolor{cyan}&&\cellcolor{cyan}\\
PCA & 0.00165 &\cellcolor{cyan} 0.00057 & 0.00191 &\cellcolor{cyan} 0.00063 & 0.00146 &\cellcolor{cyan} 0.00052\\  
&&\cellcolor{cyan}&&\cellcolor{cyan}&&\cellcolor{cyan}\\
\hline
\end{tabular}
\end{center}
\label{tab:multicol}
\end{table}

In the interest of space we will not present all 60 plots of dimension reductions.  The interested reader, however, may email the corresponding author to see those plots.

\newpage

\section{Conclusion and Future Work}

Let's conclude by giving the skeleton of our algorithm:
\begin{enumerate}
\item[1.] Given clean data ${d_i}$ of length $t$ we shift so that the minimum value is zero.
\[
d_i(\text{new}) = d_i - \min_i(d_i)
\] 

\item[2.] Center the data at $t/2$ and fit a quantum harmonic oscillator with frequency $\omega = \frac{36}{t^2}$
\[
f(\alpha) = \sum_{k=0}^{3} \alpha_k (x-t/2)^k e^{-\omega (x-t/2)^2/2}
\]

\item[3.] Divide the parameter space into $2^4$ spaces by cutting each parameter space into two pieces.  Choose the center of each cube as an initial guess.\\

\item[4.] Use the modified Levenberg-Marquardt algorithm with three iterations at each initial guess to fit the parameters. The fitted parameters are called $\alpha_*$ and the fitted oscillator $f(\alpha_*)$.\\

\item[5.] If the closest guess is much closer than the next repeat steps three and four by dividing the best guess space into smaller cubes.\\

\item[6.] Subtract the fitted oscillator from the data:
\[
d_i(\text{detrended}) = d_i(\text{new}) - f(\alpha_*)
\]

\item[7.] Fit the detrended data with a truncated Fourier series:
\[
g(\beta) = \sum_{k=0}^{1} \beta_{2k} \cos(\beta_{2k+1} x)
\]
Fit this data exactly as in steps 3,4,5. The periodic data is $g(\beta_*)$.\\

\item[8.] Subtract the periodic data to give the new aperiodic data:
\[
d_i(\text{aperiodic}) = d_i(\text{detrended}) - g(\beta_*)
\]

\item[9.] Report the leftover aperiodic data by reporting five pieces of information: The first four centralized probabilistic moments 
\[
\left\{\mu_j(d_i(\text{aperiodic}))\right\}_{j=1}^{4}
\]
And the spectral power density
\[
\sum_{i=1}^{t} d_i(\text{aperiodic})^2
\]

\item[10.] Repeat steps 1-9 for each time series.

\item[11.] Choose a spatial clustering algorithm to cluster the resultant points.

\end{enumerate}

Our work has produced a method by which we turn time series into spatial points which works quickly and accurately at the level of data clustering.  Our choice of using the quantum harmonic oscillator to pick up the general trend has the advantage of simultaneously playing the role of an autoregressive model with 3 lags and a generalized Gaussian.  Each of these models is able to pick up a wide variety of behaviors.  In particular at the level of smooth functions we can pick up a mixture of polynomials, exponentials, and Schwartz functions.  In addition, since most of the time series we consider are of limited in length (i.e. monthly for a decade or so), this allows our model to pick up logarithmic functions and fractional powers as well.  After considering these trends we then pick up periodic behaviors.  Our method also speeds up the computations dramatically as we have modified our derivative matrix to be approximately diagonal.  In particular, we add multiple rank one matrices to a diagonal matrix, and thus inverting these matrices can be done quickly via the Sherman-Morrison theorem.  The overall result is that we lose no accuracy at the level of clustering, while speeding up our original model by 2.5 orders of magnitude.  Even in time trials against the well-formed Cesium package in Python we exhibit a an increase in speed upwards up an order of magnitude in the best cases (PCA across the board), and at worst trail by roughly 9\% (t-SNE, raw in 3D).\\

In future work we wish to improve our method in several different directions.  Our first project is to normalize the data slightly further in preprocessing so that we can reduce our dimensions even further from 13 to 10.  The second major project in our scope is to change our distance measure from nonlinear least squares to dynamic time warping.  Other potential distance measures we hope to explore are dynamic quantum evolution\cite{QuantumDynamicClustering}, truncated discrete Fourier transforms and wavelet transforms.  It is our belief, based on empirical testing, that we can still improve our algorithm for speed or accuracy, but achieving both simultaneously is an extraordinary task.\\

Previously, we have produced an in-house error measurement which we call the Nousot Predictive Error.  This bases our error measurement on recent events rather than over all time.  In this work we use truncated Dirichlet series.  It is an interesting theoretical question as to whether our errors model the character of a group up to noise and how the topology of such a Lie group affects the clustering property.\\

Our future work also includes simulations in quantum computing space.  The basic task at hand is to turn our distance function into an energy landscape.  This is partly what our method does at present; the difference is the way in which we look for a global minimum.  In quantum mechanics a particle can tunnel through energy barriers which are not ``too energetic." This is the supposed approach of D-wave \cite{Dwave,Dwave2}.

\section*{Appendix: The Robustness of the Quantum Harmonic Oscillator}

Here we simply present some fitted curves which are distinct from Gaussian curves, just to show the types of trends the oscillator can fit.\\

\begin{ex}
Consider the sequence ${1,2,3,\dots, 100}$

We have the oscillator fitting using the ground state and three excitations.

\[
\includegraphics[height = 10cm,width = 30cm]{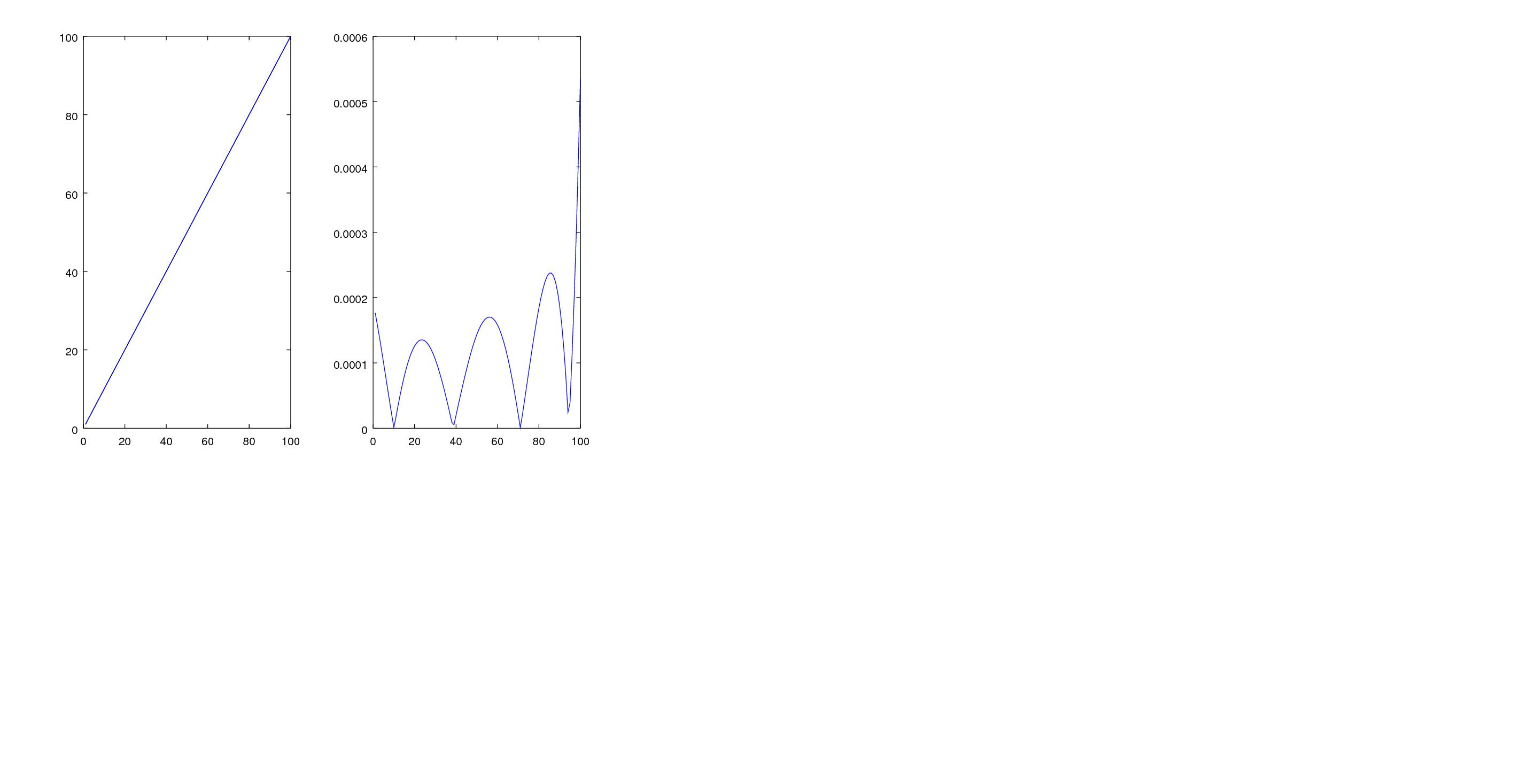}
\]

On the left we have the sequence plotted in black and the model fitted by the oscillator in blue.  On the right we have the absolute values of errors at every point.  We see a maximum error of approximately 0.0006.
The coordinates are
\[
\alpha_0 = 50, \alpha_1 = .99999,\alpha_2= 1.6257e-04, \alpha_3 = 3.2551e-06
\]

\end{ex}

\begin{ex}
Now let's consider fitting $x^2$ and $(x-a)^2$.  In this instance we'll consider $\{1^2,2^2,\dots, 100^2\}$ and then this sequence centered $\{(1-50)^2,(2-50)^2,\dots (100-50)^2\}$

Here they are

\[
\includegraphics[height = 10cm,width = 30cm]{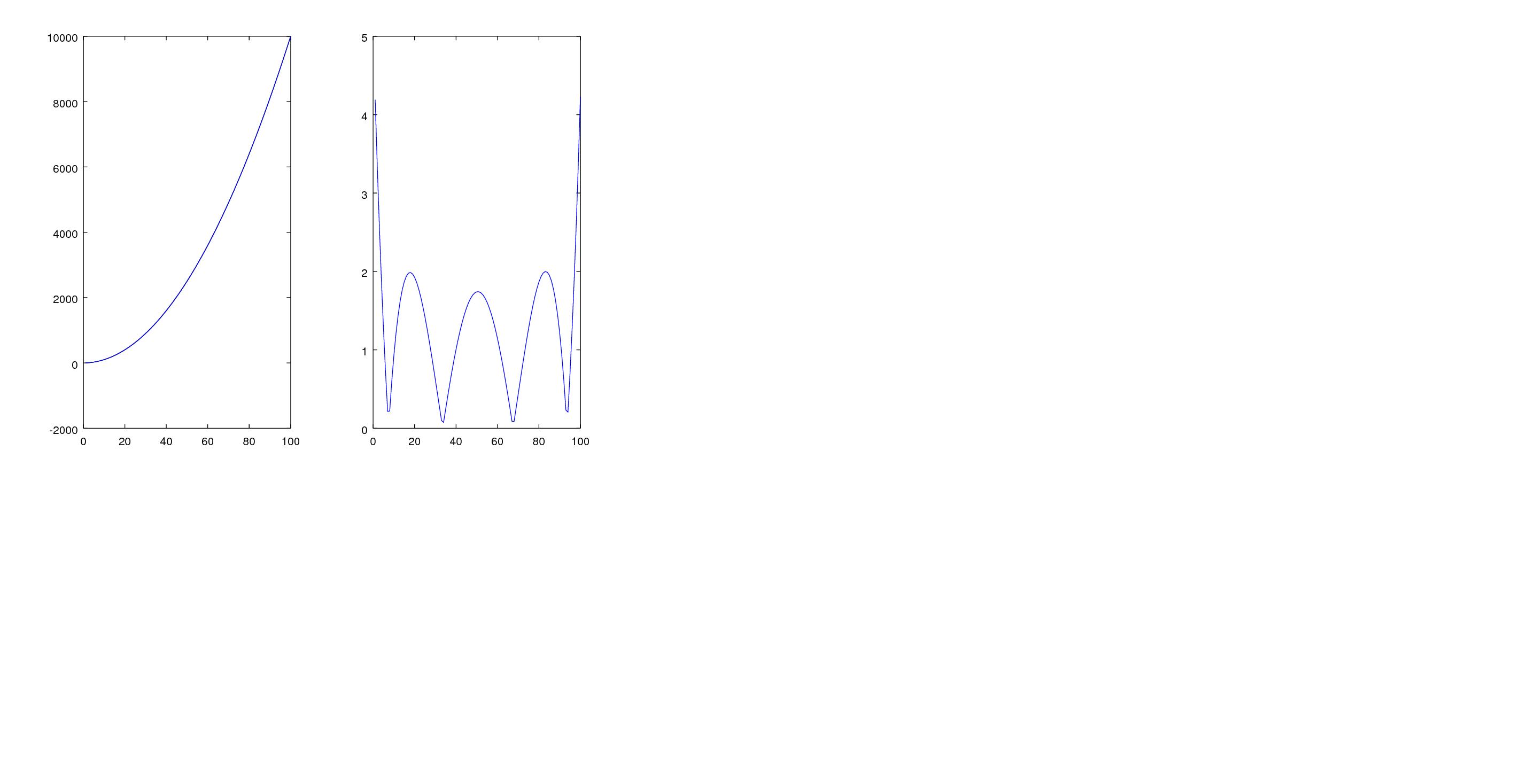}
\]

This has coordinates
\[
\alpha_0 = 2.4983e+03, \alpha_1 = 9.9992e+01,\alpha_2 =   1.0151e+00, \alpha_3 =   3.3200e-04
\]

\[
\includegraphics[height = 10cm,width = 30cm]{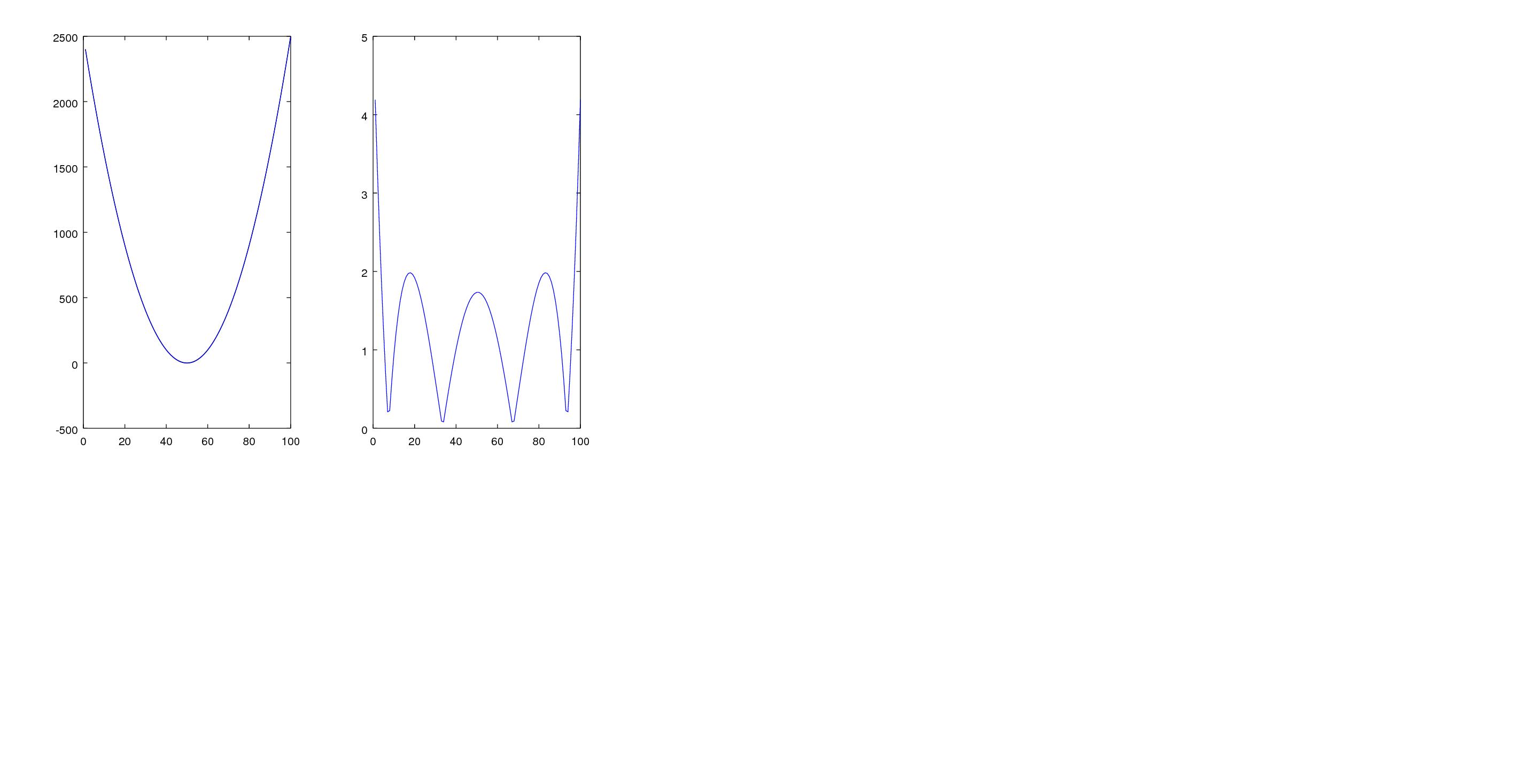}
\]

This has coordinates
\[
\alpha_0 = -1.7329e+00, \alpha_1 = -6.9565e-03, \alpha_2 =    1.0069e+00, \alpha_3 = 6.5104e-06
\]

In both cases we see a maximum absolute error of about 4, whereas the maximum values are 10000 (relative error of .0004) and 2500 (relative error of 4/2500)

The quantum harmonic oscillator can also fit with high accuracy general quadratics $ax^2+bx+c$ even where $a$ is large or negative.  These take up a lot of space to plot, but one can email the corresponding author for the code in Octave.

\end{ex}

\begin{ex}
Let's consider one final example.  This will be a skewed Gaussian.  Remember, we worked hard to center our curve.  So Let's see how well our three excitations can pick up a skewed curve.  Here, we've centered our data at 40, and our initial guess is a centered Gaussian at 50.

\[
\includegraphics[height = 10cm,width = 30cm]{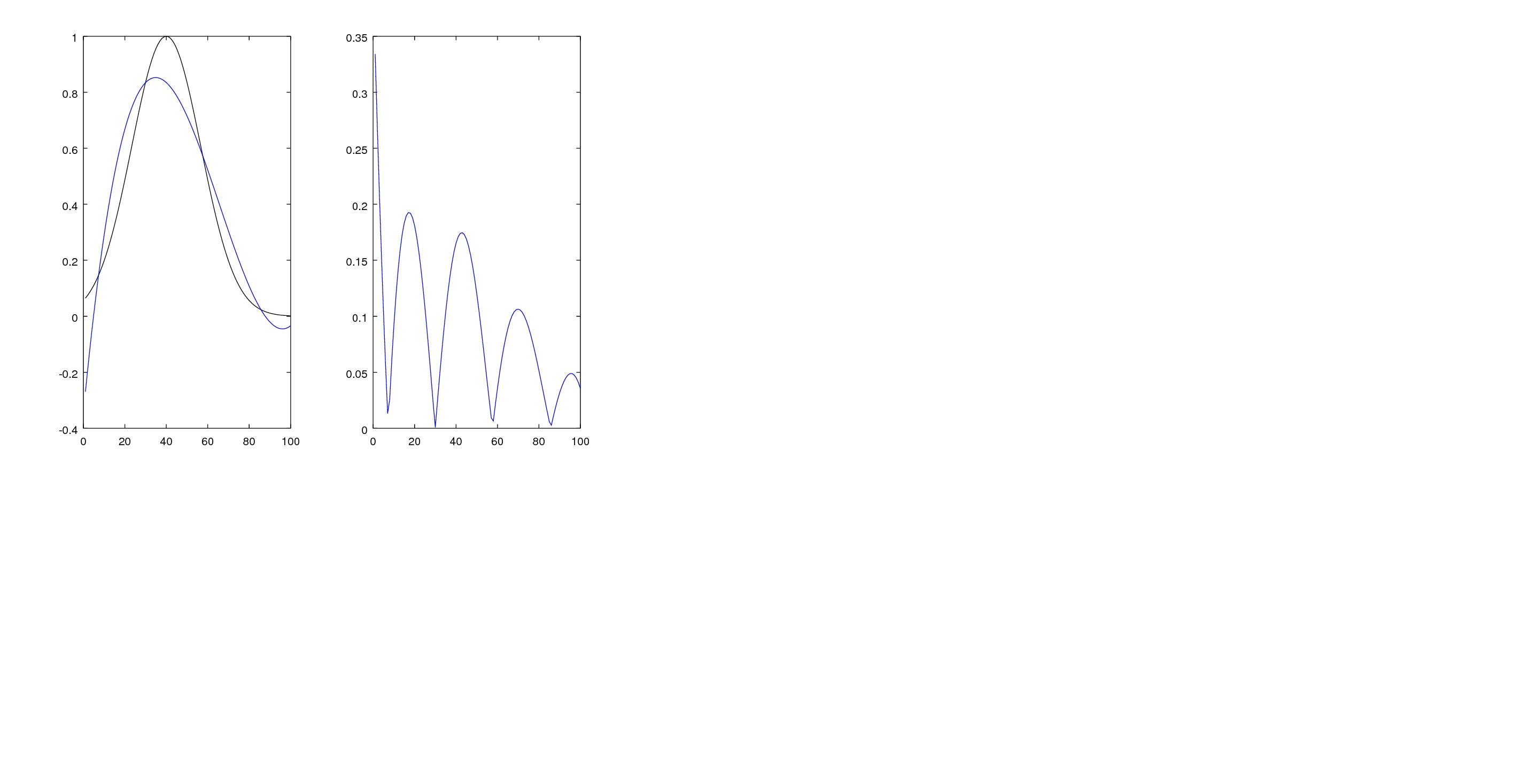}
\]

This has coordinates
\[
\alpha_0 = 7.1510e-01, \alpha_1 = -1.6399e-02, \alpha_2 =    -3.6256e-04, \alpha_3 = 7.8166e-06
\]

Here we see that the oscillator picks up the skewed nature of the data, although three excitations is not enough to get as close as we might like.  Given more energy levels we can fit this much more closely.  However, as we've discussed at length, our goal is the compute things quickly, up to accuracy in data clustering.  Looking at our errors, we see a nearly periodic function, thus our second round of coordinates would also pick this up.

\end{ex}


\begin{thebibliography}{99}
\bibitem{AlAk} \emph{A Family of Metrics for Clustering Algorithms},\\
Alexander, C., Akhmametyeva, S., arxiv.org/1707.08912

\bibitem{Cesium} \emph{cesium: Open-Source Platform for Time-Series Inference.}\\
Naul, B., van der Walt, S., Crellin-Quick, A., Bloom, J., Pérez, F. (2016). 

\bibitem{Dwave} \emph{Can Quantum Monte Carlo Simulate Quantum Annealing?}\\
Andriyash, E.,Amin, M., arxiv.org/1703.09277

\bibitem{Dwave2} \emph{Thermally Assisted Quantum Annealing of a 16-qubit Problem}\\
Dickson et el., Nature Communications volume 4, Article number: 1903 (2013)
doi:10.1038/ncomms2920


\bibitem{Evans} \emph{Partial Differential Equations}, L. Evans, Graduate Studies in Mathematics, 2nd. ed.

\bibitem{Keogh1} \emph{Dimensionality Reduction for Fast Similarity Search in Large Time Series Databases},\\
Keogh, E., Chakrabarti, K., Pazzani, M. et al. Knowledge and Information Systems (2001) 3: 263. https://doi.org/10.1007/PL00011669

\bibitem{Keogh2} \emph{On the Need for Time Series Data Mining Benchmarks: A Survey and Empirical Demonstration},\\
Keogh, E.,  Kasetty, S. Data Mining and Knowledge Discovery (2003) 7: 349. https://doi.org/10.1023/A:1024988512476


\bibitem{Keogh3} \emph{Querying and mining of time series data: experimental comparison of representations and distance measures}, \\ 
Ding, H., Trajcevski, G., Scheuermann, P. ,Wang, X., Keogh, E.\\
Proceedings of the VLDB Endowment ,Volume 1 Issue 2, August 2008 

\bibitem{Keogh4} \emph{Searching and Mining Trillions of Time Series Subsequences under Dynamic Time Warping},\\
Keogh et el., SIGKDD 2012

\bibitem{LM1}  \emph{A Method for the Solution of Certain Non-Linear Problems in Least Squares}\\
Levenberg, K.,  Quarterly of Applied Mathematics. 2, 1944.

\bibitem{LM2} \emph{An Algorithm for Least-Squares Estimation of Nonlinear Parameters}\\ 
Marquardt, D., SIAM Journal on Applied Mathematics. 11 , 1963 doi:10.1137/0111030.

\bibitem{LM3} \emph{Algorithms for the solution of the nonlinear least-squares problem}\\
Gill, P., Murray W., SIAM Journal on Numerical Analysis. 15 (5), 1978. doi:10.1137/0715063.

\bibitem{LM4} \emph{The solution of nonlinear inverse problems and the Levenberg-Marquardt method}\\
Pujol, J., Geophysics. SEG. 72 (4), 2007,  doi:10.1190/1.2732552.

\bibitem{Manifold} Scikit-learn: Machine Learning in Python, Pedregosa et al., JMLR 12, pp. 2825-2830, 2011.\\

\bibitem{QuantumDynamicClustering} \emph{Dynamic quantum clustering: a method for visual exploration of structures in data}\\
Weinstein, M., Horn, D.,	arXiv:0908.2644


\end{thebibliography}
\end{document}